# How a Vicinal Layer of Solvent Modulates the Dynamics of Proteins

**Canan Atilgan, Ayse Ozlem Aykut, Ali Rana Atilgan**

Faculty of Engineering and Natural Sciences, Sabanci University, 34956 Istanbul, Turkey,

## ABSTRACT

The dynamics of a folded protein is studied in water and glycerol at a series of temperatures below and above their respective dynamical transition. The system is modeled in two distinct states whereby the protein is decoupled from the bulk solvent at low temperatures, and communicates with it through a vicinal layer at physiological temperatures. A linear viscoelastic model elucidates the less-than-expected increase in the relaxation times observed in the backbone dynamics of the protein. The model further explains the increase in the flexibility of the protein once the transition takes place and the differences in the flexibility under the different solvent environments. Coupling between the vicinal layer and the protein fluctuations is necessary to interpret these observations. The vicinal layer is postulated to form once a threshold for the volumetric fluctuations in the protein to accommodate solvents of different sizes is reached. Compensation of entropic-energetic contributions from the protein-coupled vicinal layer quantifies the scaling of the dynamical transition temperatures in various solvents. The protein adapts different conformational routes for organizing the required coupling to a specific solvent, which is achieved by adjusting the amount of conformational jumps in the surface-group dihedrals.

## INTRODUCTION

Proteins are complex systems made up of many conformational sub-states (1), mainly determined by the folded protein structure, yet strongly influenced by environmental factors such as the solvent, temperature or pH. An understanding of the role of solvent on protein dynamics will inevitably lead to an understanding of how proteins function while responding to different environments. It will also provide clues for designing protein-solvent formulations with improved stability.

The dynamics and function of proteins were shown to be coupled to motions in the bulk solvent and the hydration shell (2). Therein, three types of protein motions were identified: (i) Those that are coupled to the dielectric fluctuations in the bulk solvent, mainly associated with the few, large-scale conformational changes such as the entrance and exit of ligands; (ii) hydration-shell coupled motions that follow the fluctuations in the hydration shell, most likely involving side-chains (3, 4) and permitting processes such as the passage of ligands inside the protein; and (iii) vibrational motions that are only coupled to the internal dynamics of the protein molecule.

Investigations on the temperature-dependent properties of folded proteins reveal that hydrated proteins show a dynamical transition above which temperature they are biologically active (5). A similar behavior has been observed in other solvents such as glycerol (6) and trehalose (7), as well as solvent mixtures such as glucose-water matrices (8), or glycerol-water mixtures (9). In our previous work we have investigated the temperature dependent thermodynamical and dynamical properties of the protein dynamical transition through molecular dynamics (MD) simulations on bovine pancreatic trypsin inhibitor in water (10, 11). We have characterized the dynamical transition using different approaches, and proposed that with the onset of the transition, an insertion of time scales occurs, due to the motions in item (ii), that operate between the slow mode motions of the protein and fast vibrational dynamics, outlined above in items (i) and (iii), respectively (11). Thus, the dynamics between these motions that are well separated in frequency become coupled, leading to a functional protein. Therein, it was suggested that those intermediate time scales were brought about by the side-chain conformational jumps.

The biological function of proteins is affected by the structural fluctuations among the conformational sub-states (12) and the solvent plays a key role in their activation (13). It has been reported that there is a correlation between protein dynamics and the thermal motion of water (14). The water molecules promote these fluctuations through a hydrogen bond network, so that the conformational fluctuations in the protein occur on the same (picosecond) time scale as the fluctuations in water (15). MD simulations on lysozyme showed that the dynamics of solvent – protein interactions along the surface control the structural relaxation of the protein as a whole (6), the effect propagating into the core of the protein through intermolecular interactions, also for the glycerol solvent. The protein, in return, has an influence on the solvent dynamics so that the solvent molecules in close proximity of the protein surface show decreased mean square fluctuations and increased relaxation time. It is known that the shell of water around the protein surface display dynamics that is distinct from the bulk. This layer – termed biological water or vicinal water – has slower relaxation behavior measured by femtosecond lasers (16) and MD simulations (17), displaying a bimodal character on the pico- and sub-picosecond time scales. Both relaxation times are slower by about one order of magnitude in the vicinal layer, the effect being most effective up to 4 Å from the surface. This may due to the hydrogen bonding or other close-range favorable interactions between the solvent molecules and the side chains on the surface of the protein; in fact, at a distance of ca. 7 Å all water molecules have bulk properties. The hydrogen bonding slows down the rotational and translational motion of solvent molecules in close proximity to protein surface (4), so that the solvent dynamics are suppressed as the solvent molecules get closer to the protein.

The motion of the solvent molecules is slow and traps the protein molecules in long lived conformations below the transition; conversely, above the transition solvent dynamics is fast enough to let the protein sample several conformations (18). MD simulations by Vitkup et al. showed that the magnitudes of the protein fluctuations are largely determined by solvent viscosity (18). As a result, high solvent viscosity plays an essential role in inhibiting the dynamics of the protein below the transition (6, 19) and protein dynamics may be suppressed by highly viscous solvents even at room temperature (19). The rate of conformational changes in myoglobin measured with nanosecond lasers hint that the dynamics is controlled by the solvent (at high solvent viscosity), the protein (at low solvent viscosity), or a crossover regime where both are effective (20). Neutron scattering analysis of xylanase in powder form, in $D_2O$ and in four two-component perdenaturated single-phase cryosolvents showed the general features of the dynamic transition behavior (21). The fast picosecond fluctuations of concentrated protein solutions were shown to follow those of the pure solvents and exhibit similar picosecond dynamic transition behavior.

All of the results outlined above indicate that the dynamics of the protein are strongly influenced by the solvent, and therefore the temperature dependence of the solvent. Although the dynamics of vicinal water has been studied in some detail, the change in the dynamics of the protein in the presence of the solvent has not been characterized. In this work, we study solvent mediated mechanisms that lead to the coupling between the fast vibrational dynamics of the protein and the slow, large scale conformational changes in the protein. We present the results of a series of extensive MD simulations on *Trichoderma reesei* Endoglucanase III in two different solvents, water and glycerol. We first characterize the common features that arise as a result of the dynamical transition in the equilibrium properties and backbone dynamics of these systems. We propose a viscoelastic model that describes the changes in the chain flexibility and the relaxation times of the backbone fluctuations with temperature and solvent. We further put forth a thermodynamical model crudely approximating the shifts in the dynamical temperature in different solvents, provided that they display similar types of interactions with proteins. In both models the main assumption is that a coupling between the vicinal solvent layer and the protein in the presence of the bulk solvent exists above the dynamical transition. The extent of coupling is shown to be manifested in the dynamics of side-chains that protrude into the solvent, the distribution of the inserted time scales being narrower in glycerol than in water. Implications of this novel approach in studying the solvent-mediated dynamics of proteins are discussed in conjunction with the stability of the protein backbone.

## SIMULATION PROTOCOL

A series of MD simulations under constant temperature – volume conditions were performed on Endoglucanase enzyme (Protein Data Bank (22) (PDB) code: 1H8V) in water and glycerol at temperatures spanning the range below and above the transition. 1H8V is a 218 amino acid protein (24.5 kDa), neutral pI, glycoside hydrolase family 12 cellulase that lacks a cellulose-binding module. It has a large substrate binding groove formed by the β sheets in which the active site residues are located (Figure 1). Active site residues Asp99, Glu116 and Glu200 from a carboxylic acid trio (23).

The NAMD package was used to model endoglucanase III, water (TIP3P) and glycerol (24). The force field parameters were used from CharmM 22 for the former two and CharmM 27 for the latter (25). The electrostatic potentials on glycerol atom surfaces were calculated using the software package GAUSSIAN 03 (26). The equilibrated glycerol box containing 400 molecules was formed and optimized with the Accelrys Materials Studio package (27); the equilibrated box of TIP3P water presented in the NAMD package is used for solvation.

The protein-solvent mixtures were formed with the VMD 1.8.5 solvent plug-in version 1.2 (28). The protein was soaked in a solvent box such that there is at least a 5 Å layer of solvent in each direction from any atom of the protein to the edge of the box. The protein-solvent mixtures were then neutralized using VMD auto-ionize plug-in version 1.2. The protein-water system contains two sodium ions and one chloride ion along with 3055 water molecules, whereas the protein-glycerol system neutralized with one sodium ion has 592 glycerol molecules. The constructed protein-water box has 12301 atoms with box dimensionality 111.8×51.3×64.2 Å. The protein-glycerol system was formed of 11452 atoms with box dimensions of 111.8×54.9×66.9 Å.

Long range electrostatic interactions were calculated using particle mesh Ewald (PME) method. The cutoff distance for non-bonded van der Waals interactions was set to 12 Å with a switching function cutoff of 10 Å. Rattle algorithm was used to fix the bond lengths to their average values. During the simulations, periodic boundary conditions were used and the equations of motion were integrated using the Verlet algorithm with a step size of 2 fs. The temperature was maintained at the desired value during the equilibration stages using temperature rescaling and during the data collection stages using Langevin dynamics. The latter treats each atom separately, balancing a small friction term with Gaussian noise to control the temperature.

The protein-solvent mixtures were minimized until the gradient tolerance was less than $10^{-2}$ with the conjugate gradients algorithm implemented in NAMD. The protein – solvent mixture was pre-equilibrated at the simulation temperature and at constant volume for 500 ps. Then the data collection run was performed at constant volume for 2 ns. The coordinate sets were saved at 2 ps intervals for subsequent analysis. The temperature range is 120 – 300 K and 130 – 385 K for the protein in water and glycerol, respectively. The temperature is incremented with steps of 10 – 15 K; the last structure saved from the lower temperature run is used as input for the equilibration run at the next temperature. A total of 34 runs were performed for these systems.

**RESULTS**

*The dynamical transition in water and glycerol*

In this study we seek to understand the effect of two different solvents – water and glycerol – on the overall thermodynamical and dynamical behavior of the system, where the term "system" refers to the protein and the solvent as a whole. We therefore first seek the dynamical transition temperature of this protein in these two solvents. Fluctuations in the thermodynamical entities such as the energy or volume are used to depict phase transitions. Their characterization is achieved by monitoring the relevant susceptibilities, e.g. the heat capacity or isothermal compressibility. Heat capacity at constant volume, $c_v$, is computed from the trajectories by using the relationship $\langle (E - \langle E \rangle)^2 \rangle / k_B T^2$ (29). In these calculations, the trajectory at each temperature is divided into 10 chunks of 200 ps length, each with 100 data points. The temperature dependence of heat capacity for the two systems is displayed in Figure 2a and the uncertainty in the data is shown with the error bars. The transition temperature is predicted to be 164 ± 4 K and 277 ± 5 K for 1h8v in water and glycerol, respectively.

It is also possible to depict the dynamical transition from an analysis of residue fluctuations. Neutron scattering measurements monitor the average fluctuations in the hydrogen atoms of the protein, $<\mathbf{u}^2>$. The theory used in interpreting those results factors out the tumbling motions undergone by the protein. In the current study and in our previous work (10, 11), following a similar treatment, we monitor the average fluctuations in $C_\alpha$ atoms of the protein, defined as $<\Delta \mathbf{R}^2>$. To achieve monitoring the internal motions only, i.e. to factor as much of the overall tumbling motion of the protein as possible, we apply the following protocol: For any 200 ps piece of the trajectory, we first make a best-fit superposition of the recorded structures to the initial structure by minimizing the root mean square deviations of the $C_\alpha$ atoms. Then the average structure, $<\mathbf{R}(T)>$, from the 100 best-fitted structures is computed. Here the brackets denote the time average. Finally, another best-fit superposition of the recorded structures to this average structure is made. Each structure of this final trajectory is denoted by $\mathbf{R}(t,T)$ and the coordinates of the $i^{\text{th}}$ residue is given by $\mathbf{R}_i(t,T)$. The fluctuation vector for a given residue $i$ at a given time $t$ from a given trajectory obtained at temperature $T$, $\Delta \mathbf{R}_i(t,T)$, is thus the difference between the position vectors for the $i^{\text{th}}$ residue of the best-fitted and the average structures,

$$\Delta \mathbf{R}_i(t,T) = \mathbf{R}_i(t,T) - \langle \mathbf{R}_i(T) \rangle \tag{1}$$

The mean square fluctuations are then obtained from an average over both time and residue index as $\langle \Delta \mathbf{R}^2 \rangle = \overline{\langle \Delta \mathbf{R}_i \cdot \Delta \mathbf{R}_i \rangle}$. Note that it is not possible to completely separate the internal motions from the external using this protocol. Nevertheless, using chunks of different sizes, 200 ps – 10 chunks, and 400 ps – five chunks, it was verified in our previous work that the nature of the transition is captured, and the transition temperature predicted does not change (11).

As a sample of protein – solvent mixture is heated starting from a temperature well below the dynamical transition, the fluctuations are expected to follow a curve that may mainly be depicted to be linear which extrapolates to zero fluctuation at absolute zero as would be expected thermodynamically. The dynamical transition is defined to occur at the temperature where a deviation from this linear behavior of $<\Delta \mathbf{R}^2>$ versus $T$ is captured. The new data again fall on a straight line, albeit the y-intercept not extrapolating to absolute zero anymore, but to a negative temperature that suggests a hysterisis in the system (30). Along with the onset of the transition, it is expected that a small fraction of the total population of particles experience barrier crossing events caused by large amplitude fluctuations of the protein conformations (31). The results are shown in figure 2b; a transition temperature of 166 ± 6 K and 267 ± 3 K for 1h8v in water and glycerol, respectively, is predicted.

Whereas the magnitude of the fluctuations experienced by the backbone $C_\alpha$ atoms, $<\Delta \mathbf{R}^2>$, gives an overall idea on the nature of the dynamical transition, the motion of the fluctuation vector carries a wealth of supplemental information. The relaxation of the $\Delta \mathbf{R}$ vector can be characterized by a relaxation function of time $C(t)$, for each temperature,

$$C(t) = \frac{\overline{\langle \Delta \mathbf{R}(0) \cdot \Delta \mathbf{R}(t) \rangle}}{\langle \Delta \mathbf{R}^2 \rangle} \tag{2}$$

where the bar and the brackets denote the average over all residues and the time average, respectively. $C(t)$ usually cannot be modeled with a single exponential decay, because there are many different contributing homogeneous processes with different relaxation times. Contributions will result in heterogeneous dynamics and assuming all contributing processes show single exponential decay, each with relaxation time $\tau_i$, can be represented by (10):

$C(t) = \sum_{i=1}^{n} a_i \exp(-t/\tau_i)$. $a_i$ is the weight with which each of these processes contribute to the observed relaxation so that $\Sigma\, a_i = 1$.

Sometimes it is possible to know the number and nature of all these processes; e.g. dielectric relaxation may usually be described by a biexponential fit. However, under many circumstances this is not possible, and furthermore all processes do not need to display Debye relaxations. One may resort to approximate the above equation by the Kohlrausch – Williams – Watts expression (32, 33): $C(t) = \exp(-t/\tau_e)^{\beta}$. Here, the stretch exponent, $\beta$, is a quantity between 0 and 1, whereas $\tau_e$ is an equivalent relaxation time for the decay of the $C(t)$ function. Note that although a value of $\beta = 1$ usually implies a single process with a single exponential decay, departure from 1 does not necessarily involve more and more complicated dynamics. In our previous study, it was shown that two equivalently contributing processes with well separated relaxation times lead to a lower $\beta$ exponent than three such processes, where a third process with an intermediate time scale is inserted between the original ones (11). The former is a "simpler" dynamics than the latter at the outset. As such, the $\beta$ exponent manifests how different contributing processes come together, leading to the observed relaxation behavior. We have previously used the quantity $\beta$ to depict the dynamical transition (10), and its variation with the temperature in water and glycerol is shown in figure 2c. The predicted transition temperature for 1h8v in water and glycerol is $182 \pm 1$ K and $280 \pm 5$ K, respectively.

The estimated transition temperatures using the three different methodologies are summarized in Table 1, along with average values. Note that the physical origin of how the predictions are made varies: Heat capacity data belong to the overall system, including both the protein and the solvent, monitoring the energy fluctuations in the system, based on information of entropic nature. $<\Delta \mathbf{R}^2>$, on the other hand, measures the amount of fluctuations experienced by the main chain $C_{\alpha}$ atoms. In this respect, it is a quantity that indirectly measures the consequences of the dynamics undergone by the system in different solvent – temperature conditions. Finally, $\beta$ is derived from the same $C_{\alpha}$ fluctuations; yet, instead of measuring the scalar amount of deviations from an average value, it stems from the dynamics of how the fluctuations vary in time. In particular, it recapitulates the combination of the many processes contributing to the relaxation of the monitored vector. In what follows, we will use how the dynamical transition surfaces in all these different phenomena to get a deeper understanding of the mechanism underlying the coupling between the dynamics of the folded protein and the surrounding solvent.

### *Solvent – protein communication leads to the dynamical transition*

Below their transition temperatures, the heat capacity of both the protein − water and the protein − glycerol systems are in the same range with values of $32.5 \pm 1.7$ and $33.4 \pm 0.5$ kcal/mol/K, respectively (Figure 2a). Therefore, prior to the onset of the dynamical transition, the energy fluctuations in the system are the same in magnitude, independent of the solvent. However, above the transition, the former system reaches a value of ca. $49.5 \pm 0.9$ kcal/mol/K whereas the latter attains a lower heat capacity of ca. $43.1 \pm 0.6$ kcal/mol/K. This observation hints that the protein and solvent begin to act as a unified and unique system at temperatures they are functional; conversely, one might conjecture that the protein behaves independently from the solvent below the transition temperature. In fact, similar observations hold for the other two system properties that we monitored in Figure 2, where the low temperature tails of the curves superimpose on each other.

Below we develop a viscoelastic model whereby we treat the protein and the solvent molecules that reside on its surface (vicinal solvent) as the system, which together are in a bath made of the rest of the solvent molecules. The equilibrium dynamics of the protein in the solvent is governed by Brownian dynamics,

$$\mathbf{Z}\Delta\dot{\mathbf{R}} + \mathbf{K}\Delta\mathbf{R} = \mathbf{F} \qquad (3)$$

where $\mathbf{F}$ is the vector holding the random forces acting on the protein and the vicinal layer, and $\Delta\mathbf{R}$ gives the mean fluctuations in the system entities, $\Delta\mathbf{R}^{\mathrm{T}} = [<\Delta\mathbf{R}>_p \ <\Delta\mathbf{R}>_v]$. Note that, in particular, $<\Delta\mathbf{R}>_p$ is given by the average of eq. 1 over all residues and $<\Delta\mathbf{R}>_v$ represents the average over all vicinal solvent molecules. The constant matrix $\mathbf{K}$ holds the spring constants of each entity, and their coupling as $\mathbf{K} = \begin{bmatrix} k_p & k_{pv} \\ k_{vp} & k_v \end{bmatrix}$ where the cross-terms are taken to be equal. $\mathbf{Z}$ describes the frictional environment, $\mathbf{Z} = \begin{bmatrix} \zeta_p & \zeta_{pv} \\ \zeta_{vp} & \zeta_v \end{bmatrix}$, with the diagonal entries representing the friction caused by the bulk solvent on the protein ($\zeta_p$) and the bulk solvent on the vicinal solvent ($\zeta_v$). The off-diagonal terms relate the friction imposed by the protein and vicinal solvent on each other, taken to be equal. Employing the fluctuation-dissipation theorem $\mathbf{FF^T} = 2k_BT\mathbf{Z}$, the time dependent auto-correlations of the fluctuations are obtained (34)

$$\langle\Delta\mathbf{R}_i(0)\cdot\Delta\mathbf{R}_i(t)\rangle = k_BT\left(e^{-t\mathbf{Z}^{-1}\mathbf{K}}\mathbf{K}^{-1}\right)_{ii} \qquad (4)$$

Where $i = 1$ for the solvent, and 2 for the vicinal layer. From eq. 4, the instantaneous elastic response of the system is given at time $t = 0$ with $<\Delta\mathbf{R}^2>_p = k_BT\ (\mathbf{K}^{-1})_{11}$. At temperatures below that of the transition $T^*$, there are no terms for the vicinal solvent, and we deal with the $1\times1$ matrix, namely a scalar, for which we recover the harmonic solution for the protein, $<\Delta\mathbf{R}^2>_p = k_BT\ /\ k_p$. At high temperatures, the average fluctuations of the protein coupled to the vicinal solvent are given by the first term on the diagonal of the solution with $<\Delta\mathbf{R}^2>_p = k_BT\ /\ k_p{'}$ where the force constant acting on the system is no longer that of the protein alone, but is a multiple of it that also depends on the elastic motion of the protein in the vicinal layer. In summary,

$$k_p^{'} = \begin{cases} k_p & T < T^* \\ k_p\left(1 - \dfrac{k_{pv}^2}{k_p k_v}\right) & T > T^* \end{cases} \qquad (5)$$

At sufficiently high temperatures, we thus get the solution with higher slopes in the fluctuation versus temperature curves, modified by the coupling between the vicinal solvent and protein. The harmonic approximation is the basis of the force constant measured by elastic neutron scattering experiments, whereby the inverse slope of the $<\Delta\mathbf{R}^2>$ versus temperature curves yields the spring constant (5, 8, 35). Similarly, from the limiting slopes of the curves in figure 2b we calculate $k_p = 4$ N/m, whereas $k_p{'} = 0.7$ and 0.9 N/m for the coupled protein-water and protein-glycerol systems, respectively.

Furthermore, since the time dependent autocorrelations are defined by eq. 2, the relaxation time for the auto-correlation of the $i^{th}$ component is obtained by integrating $\mathbf{C}(t)$ for the $i^{th}$ term:

$$\tau_i = \int_0^\infty \frac{\langle\Delta\mathbf{R}_i(0)\cdot\Delta\mathbf{R}_i(t)\rangle}{\langle\Delta\mathbf{R}_i^2\rangle}\,dt = \frac{\left(\mathbf{K}^{-1}\mathbf{Z}\mathbf{K}^{-1}\right)_{ii}}{\left(\mathbf{K}^{-1}\right)_{ii}} \qquad (6)$$

Due to the linear nature of the model, this expression predicts simple exponential decays for all the correlations. As before, below the transition temperature the matrices reduce to single term expressions due to the absence of the vicinal layer, and the average relaxation time of the protein is simply given by $\tau_p = \zeta_p / k_p$. In contrast, that above the transition may be defined to be $\tau_p' = \zeta_p' / k_p'$. The expression for the effective friction in the latter below and above the transition temperature is then:

$$\zeta_p' = \begin{cases} \zeta_p & T < T^* \\ \zeta_p \left[ 1 - \dfrac{k_{pv}^2}{k_p k_v} \left( 2 \dfrac{\zeta_{pv}/k_{pv}}{\zeta_p/k_p} - \dfrac{\zeta_v/k_v}{\zeta_p/k_p} \right) \right] & T > T^* \end{cases} \qquad (7)$$

Note that each of the ratios that appear in the bottom expression in eq. 7 is less than one. The term in square brackets gives the deviation from the protein friction coefficient. One might argue that the protein – vicinal solvent coupling will manifest itself in the stiffness and not in the damping. In that case, the second term as a whole must be much less than one. From eq. 5, we know that the $k_{pv}^2 / (k_p k_v)$ term is smaller than, but on the order of one. Moreover, the last term is negligible compared to the second because, relative to the protein, the time scale of the vicinal solvent ($\zeta_v/k_v$) is expected to be much less than that of the coupling between the protein and the solvent ($\zeta_{pv}/k_{pv}$). Thus, if there is a less-than-expected increase in the effective friction coefficient, hence the relaxation time, this will mainly be due to the further damping caused by the vicinal solvent on the protein.

In figure 3, we display the characteristic times obtained from the relaxation of the $C_\alpha$ atom fluctuations in the MD trajectories (eq. 2). These data are obtained directly from the area enclosed by the $C(t)$ curves. Although the processes recorded do not display a simple exponential decay, as conceived by the theory (eq. 6), a relative scale for characteristic times is nevertheless obtained. Since the external motions of the protein are eliminated by the structural best-fitting procedure described in the methods, the recorded times are typical of the average internal motions of the protein backbone.

As temperature is increased, and processes that involve the collectivity of larger number of atoms emerge, the relaxation time will be modified so as to yield longer times. Here we assume that the friction coefficient that arises due to the effect of bulk solvent on the protein, $\zeta_p$, remains roughly constant over the temperature range studied since protein may be considered to be a molten solid (36). The same treatment was also made by Ansari et al. in explaining the crossover between the solvent- and protein-controlled kinetics governing the conformational changes following the photodissociation of carbon monoxide in myoglobin using nanosecond laser measurements (20). Here, the fact that the relaxation times well-below the transition temperature are constant in figure 3 corroborates this assumption. We further find that the term $\zeta_{pv}$, due to the friction arising from the coupling between protein and vicinal water, is roughly independent of temperature, since the relaxation times well-above the transition are also constant. Then the leading term in eq. 9 shows that the main adjustment in $\tau'_p$ is due to the altered flexibility of the protein, namely if $\zeta_{p'} \approx \zeta_p$ then $\tau_{p'}/\tau_p = k_p/k_{p'}$. However, although the expected increase in $\zeta_p$ is by a factor of 5.7 and 4.4 for water and glycerol solvents, respectively, the observed increase in the relaxation time is only nearly by a factor of 2 in both solvents. Thus, the presence of the vicinal solvent molecules does not only cause an increase in the flexibility of the protein, but also modifies the effective friction that is felt by the protein.

### *The transition temperature is determined by the vicinal solvent layer*

In figure 3, the data are also plotted so that the temperatures of water relaxation data are shifted by a factor of ($T^*_{glycerol}/T^*_{water}$); the two curves are superimposable. In other words, identical relaxation profiles are obtained if the temperatures are scaled by a factor $T^*_{glycerol}/T^*_{water}$. The main contribution in determining the location of the transition temperature may be estimated using the solvent-protein coupling ideas developed in the previous subsection and thermodynamic arguments.

Suppose we idealize the system as having two stable states. In one, the protein is in direct contact with the bulk solvent (state A), and in the other, part of the solvent molecules organize into a vicinal solvent layer whereby the solvent molecules' motion is coupled to that of the protein, together existing in the bulk solvent (state B). These states are shown schematically in figure 4, upper panel. At temperatures exceeding the transition temperature, $T^*$, the coupled-solvent model (state B) is more stable, whereas at low temperatures the solute existing in the bulk solvent (state A) prevails.

The instantaneous change in the free energy of the system may be written as $dF = -PdV - SdT$, and under constant volume conditions, entropy at a given temperature is the slope of the $F$ vs. $T$ curve. Thus, thermodynamics specify that $F$ is a decreasing function of $T$, which at the bottom panel of figure 4, is schematically shown as a straight line. Stability requires that $F$ remains a minimum, and we expect a phase transition between states A and B at $T = T^*$, when the curve belonging to state B begins to remain below that of state A. At $T = T^*$ the free energies of the two are equal:

$$\left. \begin{array}{c} F^A = F^B \\ U^A - T^* S^A = U^B - T^* S^B \end{array} \right\} \tag{8}$$

We then write the contributions to the energy and entropy of the system, as the sum of those from the bulk solvent and the protein:

$$\left( U_p^A + U_{bulk}^A \right) - T^* \left( S_p^A + S_{bulk}^A \right) = \left( U_p^B + U_{bulk}^B \right) - T^* \left( S_p^B + S_{bulk}^B \right) \tag{9}$$

We assume that the bulk of the solvent (denoted by the subscript *bulk*) does not feel the effect of the solvent molecules that contribute to the vicinal solvent layer that becomes coupled to the protein; $U_{bulk}^A = U_{bulk}^B$ and $S_{bulk}^A = S_{bulk}^B$. On the other hand, the difference between the energetic and the entropic contributions from the protein (in A) and protein coupled system (in B) is denoted $\Delta U_{p'} = U_p^B - U_p^A$ and $\Delta S_{p'} = S_p^B - S_p^A$ so that $T^* = \Delta U_{p'} / \Delta S_{p'}$. The subscript *p′* refers to the fact that these differences take into account the interactions and microstates that develop with the organization of the vicinal solvent layer, additional to those intrinsic to the protein.

Then the ratio of the transition temperatures in two different solvents, e.g. water and glycerol as in the current study, is given by the expression

$$\frac{\left( T^* \right)_{glycerol}}{\left( T^* \right)_{water}} = \frac{\left( \Delta U_{p'} / \Delta S_{p'} \right)_{glycerol}}{\left( \Delta U_{p'} / \Delta S_{p'} \right)_{water}} \tag{10}$$

The energetic contribution will predominantly be due to interfacial interactions in the additional layer of modified solvent that is organized around the protein. If there are *n* solvent molecules in this layer, each with an accessible surface area of *a*, then $\Delta U$ $\alpha$ (*na*). This organization also brings an entropic cost that is proportional to the size of this layer, which may be approximated by (*nv*), where *v* is the volume enclosed by the accessible surface of one solvent molecule, so that $\Delta S$ $\alpha$ (*nv*). Assuming that the strength of the interactions in the interface is similar in these solvents, mainly due to the availability of –OH groups that interact favorably with the protein surface atoms along the surface of both types of solvents, the ratio of the transition temperatures may then be approximated by,

$$\frac{\left(T^*\right)_{glycerol}}{\left(T^*\right)_{water}} = \frac{\left(a/v\right)_{glycerol}}{\left(a/v\right)_{water}} \qquad (11)$$

We have computed the protein accessible surface areas of glycerol and water molecules by choosing the probe radius as 1.5 Å, an average value for the heavy atoms in the protein. Results are shown in Table 2. The ratio of the transition temperature of 1h8v in glycerol and water is 1.6, using the average values found in Table 1 and so is the prediction from eq. 11. Similarly, the dynamic transition temperature of hen egg white lysozyme in water, glycerol, and trehalose were determined to be 190, 300 and 350 K, respectively (6, 7, 37). Thus, glycerol again scales the transition temperature of lysozyme by a factor of 1.6. Note that the actual transition temperature in water is shifted by ca. 20 K due to the details of the protein structure. Similarly, trehalose scales $T^*$ by a factor of ca. 1.8 and the prediction from eq. 16 is 1.9; this equation is still applicable, since the trehalose molecule also has many –OH groups along its surface, and is involved in atomic interactions with the protein similar to those of water and glycerol. Given the crudeness of this approach with the assumptions involved, the success of the predictions points to the fact that many contributing effects cancel in energy and entropy differences of eq. 9 so that the scaling in the transition temperature is mainly due to the vicinal solvent effect.

### *Nature of solvent–protein communication through side–chain conformational transitions*

The behavior of average atomic fluctuations ($<\mathbf{u}^2>$ in neutron scattering experiments, $<\Delta\mathbf{R}^2>$ in this work) is consistent with the presence of a distribution of energy minima with successively higher barriers being crossed as the temperature is increased, the characteristic times for these events being on the picosecond scale (21). The relaxation time of the protein in both water and glycerol above their respective transition temperatures are found to be ca. 7.5 ps, hinting that the dynamics of the protein backbone is independent of the solvent. Thus, the average contribution of all processes affecting the relaxation is similar in both solvents, whereas the effective mechanisms which contribute to relaxation are different, as we discuss below.

Tarek and Tobias have attributed the onset of the dynamical transition to the relaxation of the hydrogen bond network in the water via solvent translational displacements (4). Therein, it was also shown that inhibition of these motions has been influential in suppressing protein motion, especially in the fluctuations of the side-chain atoms. Corroborating this view, in our previous work, we have shown that the communication between protein and solvent during and above the dynamical transition is manifested in the onset of conformational jumps of the torsional angles in surface group side-chains (11). These jumps were put forth as the motions that provide the intermediate time scales responsible for the communication between the slow globular protein motions, and fast local dynamics.

We study the $\chi_2$ angles of the seven Asp residues in the protein, six of which are located on the surface and one belongs to the catalytic triad (figure 1). We record the fraction of angles that display at least one conformational jump at a given temperature during the observation time frame of 2 ns (figure 5). As anticipated, the tendency of the dihedrals to sample their rotameric states increases as the temperature is raised. However, the behavior is solvent dependent, as well as the exact location of the monitored Asp residue. For example, in glycerol no jumps are recorded in any of these residues below the transition; conversely, in water occasional jumps occur, albeit not at the same residue (active site residue Asp 99 at 130 K, flexible loop residues Asp5 at 140 K, and Asp126 at 150 K).

In general, as the solvent molecules get more mobile with increasing temperature, they start to bump into the side chains, eventually causing them to make rotational jumps between their isomeric states. Moreover, the occurrence of these jumps increases at higher temperatures. Thus, inertial effects due to the type of solvent are also detrimental in the exact nature of the protein response. In figure 6, we display the time and temperature dependent torsional angle trajectories of two of the catalytic triad residues, Asp99 and Glu116, in the two solvents. Asp99, located on the basin of the catalytic site, makes conformational jumps below the transition temperature in water, whereas the same residue cannot do so below the transition temperature in glycerol. Since water has a small size, it may gain access to the catalytic site even at low temperatures, and occasionally gain enough energy to kick Asp99 so as to make it jump over the barrier separating the side-chain conformational states. Glycerol, on the other hand, cannot fit into the catalytic site due to its large size. It may gain access to the catalytic site and triggers Asp99 to make the jump only after the protein has enough flexibility to allow for the necessary volumetric fluctuations. This scenario is the same for Glu116 which is located at the entrance of the catalytic site. We further check if direct interactions between solvent molecules and side-chain heavy atoms are necessary for the conformational jumps to occur, by monitoring MD trajectories of Asp99 and Glu116 $\chi_2$ dihedrals of 1h8v obtained in vacuum at the temperatures of 130, 155, 180, 205, 230 and 255 K. No conformational transitions are recorded in a total of 12 ns runs (2 ns at each temperature).

Thus, larger solvents require a larger degree of coupling between solvent molecules and protein fluctuations in the vicinal layer.[1] In fact, the transition temperature in glycerol is probably postponed until the system gains large enough fluctuations to accommodate the larger solvent molecules (the mean square fluctuation from figure 2a is 1.9 and 2.5 $Å^2$ in water and glycerol at the point of their respective dynamical transitions). Once the vicinal layer forms with the establishment of the necessary communication between the solvent and solute, an even larger amount of fluctuations is introduced due to the processes that now become accessible on the potential energy surface leading to a marked increase in the flexibility. The specific interactions between solvent and solute may be due to hydrogen bonds, as emphasized in previous studies (4, 6, 39), as well as other favorable interactions between polar atoms. In fact, we have monitored solvent molecules near the catalytic site residues, and find that a close distance is maintained between the nearest oxygen atom of the solvent and the side-group heavy atoms; a collinear hydrogen bonding angle is not necessarily retained between the donor and acceptor groups.

The extent of coupling between backbone positional fluctuations and conformational jumps in the torsional angles of chain molecules markedly affect the dynamics of the system (40). In general, the conformational entropy of the system is distributed between the backbone and side-chains. The force constant of the protein in glycerol is larger than that in water (eq. 5 and figure 2b) indicating a less flexible backbone once the system has gone through the dynamical transition. The protein in glycerol also exhibits a somewhat lower ability to make side-group torsional jumps. Furthermore, the stretch exponent, $\beta$, provides information about the distribution of time scales contributing to the relaxation of the protein in the solvent. In the protein-water system ($\beta$ = ~0.5 in figure 2c), as opposed to the protein-glycerol system ($\beta$ = ~0.4), a wider distribution of time scales, and therefore a larger variety of processes contributing to relaxation, is implied.

---

[1] Note that the effect of solvent on the size or shape of the protein is a matter of discussion in the literature (see, e.g., ref. (38) and references cited therein; Ansari et al. (20) report that increasing glycerol concentration in solvents of glycerol-water mixtures has no perceptible effect on the size or shape of myoglobin). Here, the size of the protein remains essentially the same in both solvents, showing ca. 1% difference when averaged over all frames collected throughout the simulations which total to more than 40 ns in each solvent.

**CONCLUSIONS**

Using thermodynamical and viscoelastic arguments, we explain the nature of the protein dynamical transition and its dependence on different solvents by studying the equilibrium dynamics of the protein 1h8v in water and glycerol at a variety of temperatures, all below the unfolding temperature. Mean positional fluctuations, a scalar which may be thought of as the fluctuations of the protein mass center, display a slope change at a critical temperature. When the critical temperature is reached, the vicinal solvent, which is treated as another unified mass coupled with the mass center of the protein, not only shifts the mass center, but also its fluctuations by weakening the effective spring constant (figure 2b and eq. 5). We, metaphorically, state that, below the critical temperature, the mass center of the protein is decoupled from the solvent; i.e. vicinal solvent is indifferent from the bulk below the critical temperature. We depict this change in the states as the protein dynamical transition, sometimes referred to as the protein glass transition.

Relaxation time versus temperature graph also shows a state change at the critical temperature (figure 3). This is the relaxation time of the mass center's positional fluctuation, requiring damping or friction operating on the mass center of the protein (eq. 7). Below the critical temperature, since the mass center of the protein independently communicates with bulk solvent, the latter impinges retardation on the protein referred to as $\zeta_p$. Above the critical temperature, a similar frictional effect, $\zeta_v$, from the bulk solvent also influences the vicinal layer. Furthermore, since an interaction is also switched on between the masses, symbolized by $k_{pv}$, $\zeta_{pv}$ is induced concomitantly. This is a further change in the entropy due to the additional excluded volume effects that bulk solvent creates simultaneously on vicinal solvent and the protein.

The existence of $\zeta_{pv}$ can be qualitatively observed on the relaxation times. It introduces another time scale to the problem in the following way: $\zeta_p/k_p$ and $\zeta_v/k_v$ are the two scales of the protein and the vicinal layer, respectively, independent of each other. Their existence is due to the decoupled interactions of protein with the bulk and the vicinal layer with the bulk. Below the transition only the former exists, and the additional conformational motions are not induced, justified by the low number of conformational transitions observed in the MD simulations. Right after the transition, another time scale is introduced mainly due to $\zeta_{pv}$, which modifies the relaxation time given approximately by $\tau_p \approx \zeta_p/k'_p$, if only the alteration of the $\zeta_p/k_p$ due to energetic interaction is considered. Yet, the concomitant entropic term is not there until $\zeta_{pv}$ is also considered and $\tau_p$ is further modified (eq. 7).

Thus, interaction energy and entropy are due to $k_{pv}$, which alter the mean positional fluctuation slope, and $\zeta_{pv}$ which affects the relaxation times by introducing another independent time scale. This effect is validated by the additional number of conformational transitions recorded from the MD simulations. Therefore, a unified Brownian dynamics with different stiffness and friction terms outlines all the observations made for the folded protein at a large variety of temperatures and in different solvents. The competition between the entropic and energetic factors that arise due to the interactions between the vicinal layer and the protein further explain the alteration in the value of the transition temperature in different solvents if they impinge similar types of interactions on the protein surface.

The details of the phenomena observed at the level of molecular detail are manifested in data reflecting the collective behavior of the system. The heat capacity of the system above the transition is less for the protein-glycerol than the protein-water system (figure 2a). The smaller water molecules are less restricted to move around the side chains, compared to glycerol molecules, leading to a more entropic state for the protein-water system. This is also in agreement with our findings for residue fluctuations (figure 2b), further supported with the torsional angle trajectories of figure 6. For the protein-water system, even before the onset of the transition, the side chains have a chance to make jumps due to the high entropic contribution of water, providing communication between different sub-state conformations. Apart from the monitoring of specific dihedrals, figure 5 shows that in general the jump rates increase with the increased mobility of the protein, more so in water than in glycerol.

Side-chain conformational transitions relieve excess energy that is stored in the system. If this relaxation pathway does not exist, the protein backbone torsional angles would be forced to make similar jumps. However, such jumps are known to require cooperativity between closely located dihedral angles along the backbone of chain molecules (41, 42), so as to minimize the work done against the frictional environment during the displacement of the attached atoms (43). In proteins, these conformational transitions are prohibitive in that they lead to unfolding of the chain. In fact, we have monitored the trajectories at temperatures close to unfolding, and find that these are earmarked by attempted, but short-lived, conformational jumps on the backbone (see, e.g. the 310 K $\psi$ trajectory in figure 4 of ref.(11)). The vicinal solvent layer closely interacts with the surface groups, and provides an alternate route for the system to spend the accumulated energy while maintaining the protein with enough flexibility to perform its function.

The average relaxation time of backbone fluctuations measured for the protein in both solvents and at all temperatures above the transition temperature is the same, in spite of the different distribution of conformational routes employed in different environments (as measured by the stretch exponent $\beta$, figure 2c). Thus, the functional protein operates at a narrow time scale, and organizes its environment for achieving this operational state.

**Acknowledgements**

We thank Dr. A. S. Özen for her help in parameterizing glycerol. C.A. acknowledges support by the Turkish Academy of Sciences in the framework of the Young Scientist Award Program (CB/TÜBA-GEBIP/2004-3).

**Table 1.** Protein Transition Temperatures Obtained by Various Approaches

|  | Protein-water system | Protein-glycerol system |
|---|---|---|
| Heat capacity, $c_v$ | $164 \pm 4$ K | $277 \pm 5$ K |
| Residue Fluctuations, $<\Delta\mathbf{R}^{2>}$ | $166 \pm 6$ K | $267 \pm 3$ K |
| Stretch exponent, $\beta$ | $182 \pm 1$ K | $280 \pm 5$ K |
| Average | $171 \pm 11$ K | $275 \pm 13$ K |

**Table 2.** Accessible surface area and volume of solvent molecules

| Molecule | Surface area, $a$ (Å$^2$) | Volume, $v$ (Å$^3$) | $a/v$ (Å$^{-1}$)[†] |
|---|---|---|---|
| water | 31 | 16 | 1.9 |
| glycerol | 101 | 81 | 1.2 |
| trehalose[‡] | $268\pm1$ | $271\pm6$ | 1 |

[†]The $a/v$ ratios are the same for probe sizes ranging from $1.4 - 1.6$ Å; results displayed here are for probe radius of 1.5 Å.

[‡]Values averaged over different conformers around the torsional angle connecting the rings

**Figure Captions:**

***Figure 1:*** The three-dimensional structure of the protein *Trichoderma reesei* Endoglucanase III (PDB code: 1H8V). The three catalytic residues are highlighted in color – ASP 99 (orange), GLU 116 (red), GLU 200 (orange). The rest of the Asp residues studied are also shown in blue.

***Figure 2:*** Temperature dependence of system properties monitored for detecting the dynamical transition: a) System heat capacity; b) Thermal fluctuations averaged over all residues; c) The stretch exponent, *β*. Results from protein – water and protein – glycerol systems are shown with open and filled circles, respectively. The best fitting lines in (a) and (c) are to Boltzmann sigmoidal functions; see ref. (11) for details.

***Figure 3:*** The $C_\alpha$ relaxation times as a function of temperature on a double logarithmic scale. Open circles: protein-water; filled circles: protein-glycerol systems. The dotted curve shows the data for the protein – water system with the temperatures shifted by a factor of $T^*_{glycerol}/T^*_{water}$ (eq. 11).

***Figure 4:*** Scheme depicting the two-state model of vicinal water organization.

***Figure 5:*** The fraction of the $\chi_2$ torsional angles that display at least one jump between the conformational states at a given temperature for the seven Asp residues in the studied protein. Results from protein – water and protein – glycerol systems are shown with open and filled circles, respectively.

***Figure 6:*** Torsional angle trajectories of selected catalytic residues in different solvents over the whole temperature range studied. a) GLU 116 in water, b) ASP 99 in water, c) GLU 116 in glycerol, d) ASP 99 in glycerol. Note that the whole set of the MD simulations may be regarded as a heating procedure: The final structure from a simulation at a given temperature is the initial structure of the simulation at the next temperature. The trajectories resulting from the 0.5 ns equilibration periods (assumed to be transients between the structures equilibrated at the lower and higher temperatures) are not shown.

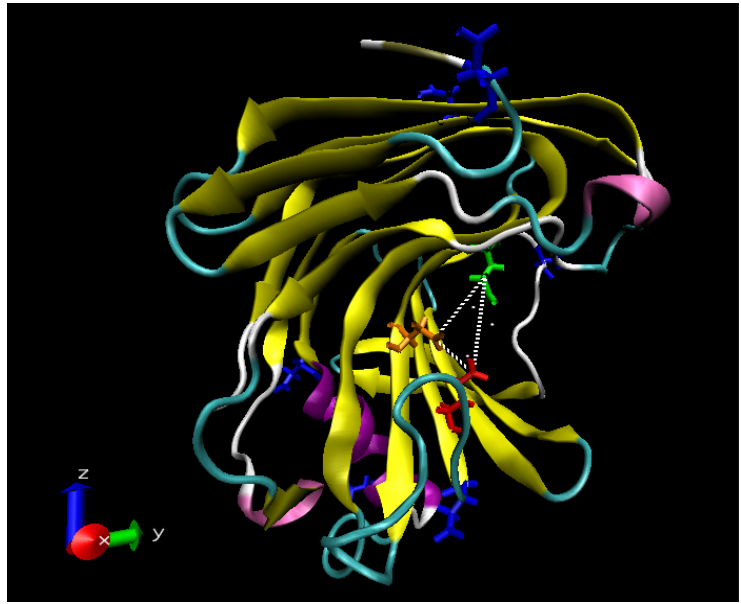

Figure 1

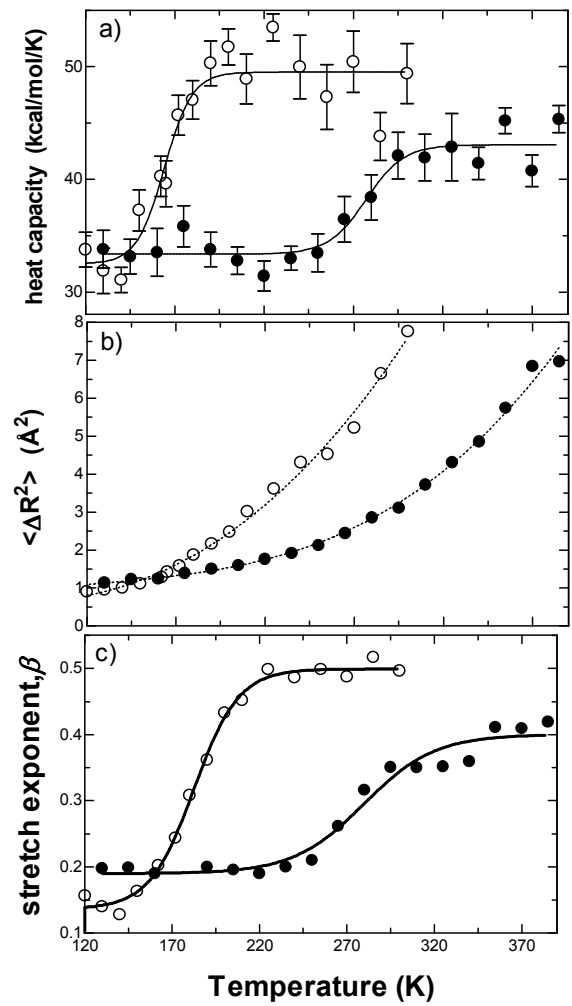

Figure 2

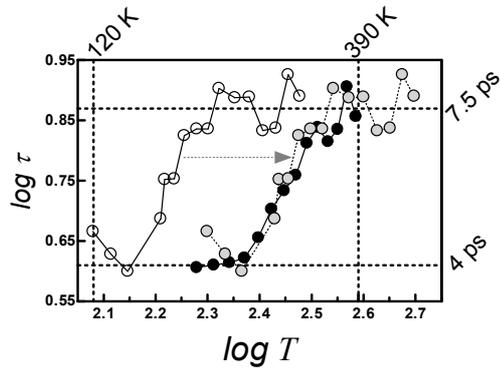

Figure 3

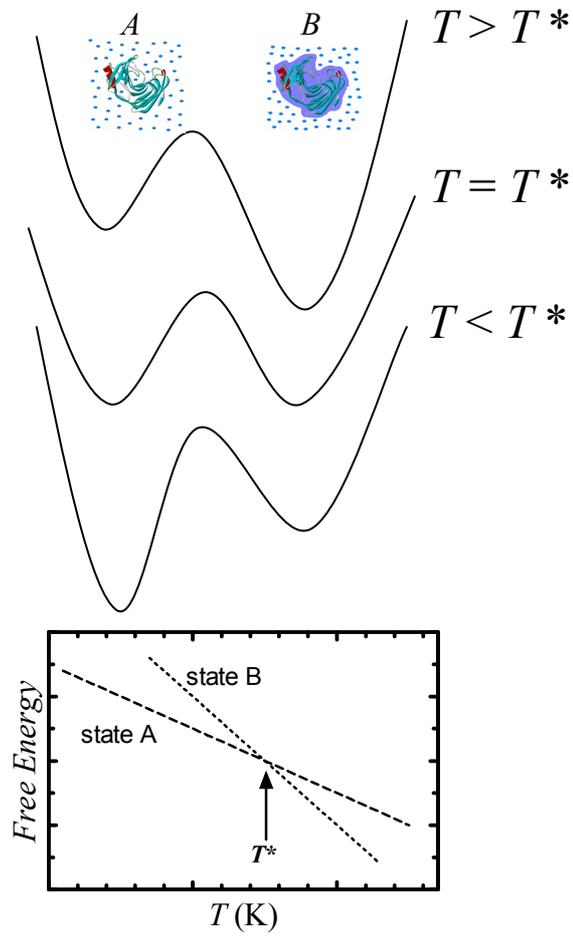

Figure 4

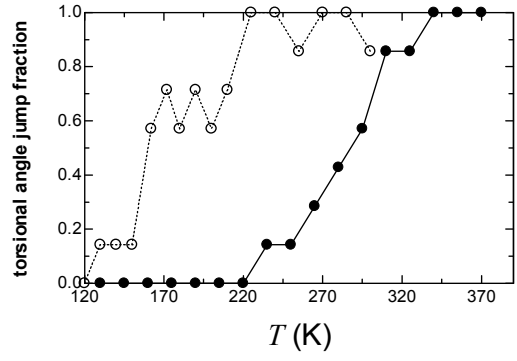

Figure 5

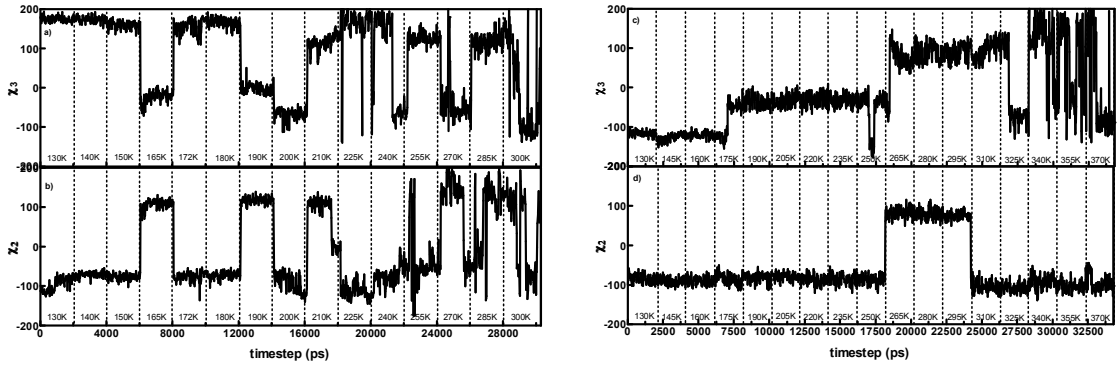

Figure 6